\documentclass{iopart}

\begin{document}
\jl{1}

\letter {On the controversy over the stochastic density functional equations}

\author {H Frusawa\footnote[1]{e-mail:
furu@exp.t.u-tokyo.ac.jp} and R Hayakawa}

\address{Department of Applied Physics, University of Tokyo,\\Bunkyo-ku, Tokyo 113-8656, Japan}

\begin{abstract}
This letter aims at justifying the stochastic equations in terms of the number density variable, which are still controversial, via complementing Dean's approach [Dean D S 1996 {\itshape J. Phys. A} {\bf 29} L613]. Our course is twofold: First, we demonstrate that standard manipulations straightforwardly transform the stochastic equation of density {\itshape operator}, derived by Dean, to the Fokker-Planck equation for the (c-number) density distribution functional $P(\{\rho\},t)$. Moreover, we verify the associated static solution of $P(\{\rho\},t)$ with the help of the conditional grand canonical partition function. 
\end{abstract}
\section{Introduction}
In supercooled liquids, due to the dense packing and strong correlation of the constituent particles, the nonvibrational diffusive motions take much more time than collisions. In other words, the momentum and the energy flow much more quickly via collisions through the system than the slowly decaying number density. Recently there has been a considerable effort to describe such slow dynamics in liquids \cite{review}. We shall in particular discuss one of the approaches, the following stochastic equations in terms of the number density field $\rho(x,t)$ [2-5]:
\begin{eqnarray}
\label{Lanresult}
\frac{\partial\rho(\,x,t)}{\partial t}=\left.\nabla\cdot L[\,\rho(x,t)\,]\,\nabla\frac{\delta H(\{\rho\})}{\delta\rho}\right|_{\rho(x,t)}+\xi(x,t),
\end{eqnarray}
or its equivalent, i.e. the Fokker-Planck equation for the probability distribution functional $P(\{\rho\},t)$:
\begin{equation}
\label{Fokkresult}
\fl\frac{\partial P(\{\rho\},t)}{\partial t}=-\int dx\,\frac{\delta}{\delta\rho(x)}\nabla\cdot L[\,\rho(x)\,]\nabla\,\bigl[T\frac{\delta}{\delta\rho(x)}+\frac{\delta H(\{\rho\})}{\delta\rho(x)}\bigr]\,P(\{\rho\},t).
\end{equation}
In equations \eref{Lanresult} and \eref{Fokkresult}, the Hamiltonian $H$ is of the free energy functional form as
\begin{equation}
\label{hamiltonian}
H(\{\rho\})=\frac{1}{2}\int\,dxdy \,\rho(x)V(x-y)\rho(y)+T\int\,dx \,\rho(x)\log\rho(x),
\end{equation}
$L[\,\rho(x)\,]$ is the kinetic coefficient written as $L[\,\rho(x)\,]=\rho(x,t)\Gamma$ with $\Gamma$ being the mobility of particles, and $\xi$ is the divergence of a random force and its correlation function is given by
\begin{eqnarray}
\label{noiseresult}
\langle\xi(x,t)\xi(y,t')\rangle=2T\nabla_x\cdot L[\,\rho(x)\,]\nabla_y\delta(x-y)\,\delta(t-t').
\end{eqnarray}
For the explicit representation of this averaging, see equation \eref{aveprocedure} below.

These equations have attractive features for studying the slow dynamics: One is the physically clear incorporation of the thermally activated hopping processes via the last random term $\xi$ on the right hand side (rhs) of equation \eref{Lanresult}. Also of interest is the density dependence of the kinetic coefficient, $L\propto\rho$, that produces the nonlinear term of dynamic origin \cite{nonlinear} and implies the relevance of these equations to describing dynamical heterogeneity \cite{onuki}.

Nevertheless the stochasticity for the number density variable is still controversial. To see this, let us first mention one of some attempts [2-4], presented by Dean \cite{Dean}, to justify the above stochastic equations \eref{Lanresult} and \eref{Fokkresult}: Consider here a canonical system of $N$ particles interacting via a pairwise potential $V(x)$ and surrounded by a thermal white noise heat bath. The $i$th particle then obeys the Langevin equation,
\begin{equation}
\label{Lanstart}
\frac{dX_i(t)}{dt}=-\Gamma\,\,\sum_{j=1}^{N}\nabla_iV[X_i(t)-X_j(t)]+\eta_i(t),
\end{equation}
where the components of the noise $\eta_i^{\alpha}$ are taken to be uncorrelated as $\langle\eta_i^{\alpha}(t)\eta_i^{\beta}(t)\rangle=2T\Gamma\delta_{ij}\delta^{\alpha\beta}\delta(t-t')$. Dean shows, using the Ito prescription for the change of variables and summing over the $i$, that equation \eref{Lanstart} is transformed to the equation of the density operator, $\hat{\rho}(x,t)=\sum_i\hat{\rho}_i(x,t)=\sum_i\delta[x-X_i(t)]$:
\begin{eqnarray}
\label{rhohat}
\eqalign{\frac{\partial \hat{\rho}(x,t)}{\partial t}=\nabla\cdot\hat{\rho}(x,t)\int dy\,\hat{\rho}(y,t)\nabla V(x-y)+T\nabla^2 \hat{\rho}(x,t)+\hat{\xi}(x,t)\\
\hat{\xi}(x,t)=-\sum_i\nabla\cdot[\hat{\rho}_i(x,t)\eta_i(t)].}
\end{eqnarray}
Since one finds
\begin{eqnarray}
\label{aveprocedure}
\langle\hat{\xi}(x,t)\hat{\xi}(y,t)\rangle&\equiv\int d\eta_i\,\hat{\xi}(x,t)\hat{\xi}(y,t)\,\exp\left(-\int dt\, \frac{\eta_i^2}{4T\,\Gamma}\right)\\
\label{corroperator}
&=2T\nabla_x\cdot L[\,\hat{\rho}(x)\,]\nabla_x\delta(t-t'),
\end{eqnarray}
equation \eref{Lanresult} is verified so long as the operator $\hat{\rho}$ reads $\rho$ of c-number.

As expected, though, Marconi and Tarazona (MT) \cite{MT} subsequently object to the last supposition: they claim that $\rho$ is to be defined by averaging $\hat{\rho}$ over the noise as $\rho_{av}=\langle\hat{\rho}\rangle$, where the subscript $av$ is appended for emphasizing the procedure. As a consequence dynamical density functional equation becomes deterministic:
\begin{equation}
\label{MT}
\frac{\partial\rho_{av}(\,x,t)}{\partial t}=\nabla\cdot\int dy\,\langle\hat{\rho}(x,t)\hat{\rho}(y,t)\rangle\nabla V(x-y)+T\nabla^2 \rho_{av}(x,t),
\end{equation}
whereby the Boltzmann distribution of number density is assured as time-independent solution \cite{MT,BagMuna}.

To settle such controversy over the stochastic density functional equations \eref{Lanresult} and \eref{Fokkresult}, this letter aims at complementing Dean's argument from \eref{Lanstart} to \eref{corroperator} so that the above criticism by MT may become invalid. Our strategy is twofold: First, in the next section, we demonstrate that standard manipulations \cite{Justin} transform equation \eref{rhohat} of the density operator to the Fokker-Planck equation \eref{Fokkresult}. Moreover we verify in section 3, with the help of the conditional grand canonical partition function, the static solution $P_0(\{\rho\})$ of \eref{Fokkresult}:
\begin{equation}
\label{static}
P_0(\{\rho\})\propto\exp(-\beta H),
\end{equation}
where $\beta =T^{-1}$. In the final section, to clarify the connection between the stochastic and the deterministic equation, we confirm using the WKB-like approximation \cite{WKB} to the Fokker-Planck equation \eref{Fokkresult} that the noise-averaged deterministic equation \eref{MT} corresponds to that for the saddle-point path of $P(\{\rho\},t)$; this reveals that MT's argument produces only the mean-field equation and not the first member of the BBGKY hierarchy \cite{BBGKY} including two-point equal-time correlation function.

\section{From equation \eref{rhohat} to the Fokker-Planck equation \eref{Fokkresult}}
Turning our attention to the functional space, we immediately find that the density operator $\hat{\rho}$ may be directly mapped to the distribution functional $P(\{\rho\},t)$ as
\begin{equation}
\label{average}
P(\{\rho\},t)=\left\langle\>\prod_x\delta[\,\hat{\rho}(x,t)-\rho(x)\,]\>\right\rangle,
\end{equation}
not via the averaging $\rho_{av}=\langle\hat{\rho}\rangle$; essentially only this definition has dissolved the MT's critique.

Let us then exhibit below that $P(\{\rho\},t)$ with equation \eref{rhohat} satisfies the Fokker-Planck equation \eref{Fokkresult}. We first differentiate \eref{average} with respect to time, and obtain
\begin{eqnarray}
\label{Fokkderive1}
\fl\frac{\partial P(\{\rho\},t)}{\partial t}=\int dx\,\left\langle\>\frac{\partial\hat{\rho}(x,t)}{\partial t}\frac{\delta}{\delta\hat{\rho}(x,t)}\delta[\,\hat{\rho}(x,t)-\rho(x)\,]\prod_{y\neq x} \delta[\,\hat{\rho}(y,t)-\rho(y)\,]\right\rangle\nonumber\\
\label{Fokkderive2}
\eqalign{\fl\qquad\qquad=\int dx\,\left\langle\>\left[\nabla\cdot\hat{\rho}(x,t)\int dy\,\hat{\rho}(y,t)\nabla V(x-y)+T\,\nabla^2 \hat{\rho}(x,t)+\hat{\xi}(x,t)\right]\right.\\
\qquad\qquad\left.\times\frac{\delta}{\delta\hat{\rho}(x,t)}\delta[\,\hat{\rho}(x,t)-\rho(x)\,]\prod_{y\neq x} \delta[\,\hat{\rho}(y,t)-\rho(y)\,]\>\right\rangle.}
\end{eqnarray}
We may replace $\hat{\rho}$ by $\rho$ using the $\delta$--function, and hence equation \eref{Fokkderive2} reads
\begin{eqnarray}
\label{Fokkderive3}
\eqalign{\fl\frac{\partial P(\{\rho\},t)}{\partial t}=-\int dx\,\frac{\delta}{\delta\rho(x)}\left[\>\nabla\cdot\rho(x)\int dy\,\rho(y)\nabla V(x-y)+T\,\nabla^2 \rho(x)\right]P(\{\rho\},t)\\
\qquad\qquad-\int dx\,\frac{\delta}{\delta\rho(x)}\left\langle\,\hat{\xi}(x,t)\,\prod_{x}\delta[\,\hat{\rho}(x,t)-\rho(x)\,]\,\right\rangle.}
\end{eqnarray}
With use of the identity for an arbitrary function $F(\{\eta_i^{\alpha}\})$, $\langle\, F(\{\eta_i^{\alpha}\})\eta_i^{\alpha}(t)\,\rangle=2T\Gamma\left\langle\,\delta F(\{\eta_i^{\alpha}\})/\delta\eta_i^{\alpha}(t)\,\right\rangle$, the last bracket term on the rhs of \eref{Fokkderive3} is further transformed to
\begin{eqnarray}
\fl\left\langle\,\hat{\xi}(x,t)\,\prod_{x}\delta[\,\hat{\rho}(x,t)-\rho(x)\,]\,\right\rangle=-2T\,\Gamma\left\langle\sum_{i,\alpha}\frac{\partial\hat{\rho}_i(x,t)}{\partial x^{\alpha}}\,\frac{\delta\hat{\rho}(y,t)}{\delta\eta^{\alpha}_i}\frac{\delta}{\delta\hat{\rho}(y,t)}\prod_{y}\delta[\,\hat{\rho}-\rho\,]\right\rangle\nonumber\\
\label{math1}
\qquad=T\,\Gamma\left\langle\sum_{i}\nabla_x\cdot\nabla_y[\,\hat{\rho}_i(x,t)\hat{\rho}_i(y,t)\,]\frac{\delta}{\delta\hat{\rho}(y,t)}\prod_{y}\delta[\,\hat{\rho}-\rho\,]\right\rangle,
\end{eqnarray}
where the superscript $\alpha$ denotes the component of $x$ and $\eta_i$, and use has been made of
\begin{equation}
\label{math2}
\frac{\delta\hat{\rho}_i(y,t)}{\delta\eta_i^{\alpha}}\longrightarrow\,-\frac{1}{2}\frac{\partial \hat{\rho}_i(y,t)}{\partial y_i^{\alpha}}
\end{equation}
that is obtained from standard mathematical manipulation of the discretized Langevin equation \cite{Justin}. Also, noting that relation $\hat{\rho}_i(x,t)\hat{\rho}_i(y,t)=\delta(x-y)\rho_i(x,t)$ gives
\begin{equation}
\label{math3}
\nabla_x\cdot\nabla_y[\,\hat{\rho}_i(x,t)\hat{\rho}_i(y,t)\,]=-\nabla_x\cdot\hat{\rho}_i(x,t)\nabla_x\delta(x-y)
\end{equation}
and replacing $\hat{\rho}$ by $\rho$ as before, the bracket term finally reads
\begin{equation}
\label{lastterm}
\left\langle\,\hat{\xi}(x,t)\,\prod_{x}\delta[\,\hat{\rho}(x,t)-\rho(x)\,]\,\right\rangle=T\,\Gamma\,\nabla\cdot\rho(x,t)\nabla \frac{\delta P(\{\rho\},t)}{\delta\rho(x,t)}.
\end{equation}
Equation \eref{Fokkderive3} with this is none other than the Fokker-Planck equation \eref{Fokkresult}.

Thus it has been demonstrated that the stochastic equation of density operator \eref{rhohat} leads to the Fokker-Planck equation \eref{Fokkresult} for the (c-number) density distribution functional [or equation \eref{Lanresult}].


\section{Verification of the equilibrium distribution functional \eref{static}}
We arrive at a time-independent solution of the Fokker-Planck equation in the large time limit: $P_0(\{\rho\})=\lim_{t\to\infty}P(\{\rho\},t)$. Therefore it is plausible to suppose that the noise averaging in calculating $P_0$ becomes equivalent to the configurational one in equilibrium:
  \begin{equation}
  \label{configaverage}
  \fl \,P_0(\{\rho\})\propto\frac{1}{N!}\prod_{i=1}^N\int\,dX_i
  \prod_x\delta[\,\hat{\rho}(x,t)-\rho(x)\,]\,\exp[-\beta\sum_{i,j}V(X_i-X_j)].
  \end{equation}
The problem is then how to derive expression \eref{static} from the above conditional partition function.

Let us once move to the grand canonical system where we are to consider
\begin{equation}
\label{grand}
P_0^{\Xi}=\sum_{N=0}^{\infty} \,P_0\lambda^N,
\end{equation}
with $\beta\mu=\ln\lambda$ being chemical potential. Introducing the auxiliary field $\psi$ as $\delta[\,\hat{\rho}(x)-\rho\,]=\int{d}\psi\;\exp\>[\,i\psi(\hat{\rho}-\rho)\,]$, the configurational representation of $P_0^{\Xi}$ given by \eref{configaverage} and \eref{grand} reads
  \begin{equation}
  \label{func1}
  \fl P_0^{\Xi}\propto\int{D}\psi\;\exp\>\left[\,-\beta\int{d}x\,{d}y\;\frac{1}{2}
  \,\rho(x)\,V(x-y)\,\rho(y)+i\rho(x)\,\psi(x)-e^{i\psi(x)+\mu}\right],
  \end{equation}
where $D\psi$ is formally defined as $\prod_xd\psi(x)$. Since there is no contribution to $P_0^{\Xi}$ of principal quadratic fluctuation of the auxiliary field $\psi$ around the saddle point path $\psi_{sp}$ as shown elsewhere \cite{frusawa}, Gaussian approximation for $\psi$ reduces the functional integral form \eref{func1} to
  \begin{equation}
  \label{func2}
  P_0^{\Xi}\propto\exp\>\left[\,-\beta H+\beta\int{d}x\;\rho(x)+\mu\rho(x)\right],
  \end{equation}
as found from substituting $\rho=e^{i\psi_{sp}+\mu}$ into \eref{func1}.

For returning to the canonical system, we have only to perform the following contour integral,
\begin{equation}
\label{gtoc}
P_0=\frac{1}{2\pi i}\oint d\lambda \frac{P_0^{\Xi}}{\lambda^{N+1}},
\end{equation}
where $\lambda$ is now a complex variable. This relation with use of the Cauchy's integral theorem gives back the canonical form:
  \begin{eqnarray}
  \label{canfunc}
  P_0\propto\exp\left[-\beta H+\beta\int dx\rho(x)\right]
  \frac{1}{2\pi i}\oint d\lambda\; \frac{1}{\lambda^{1+[N-\int{dx}\rho(x)]}}
  \nonumber\\
  \qquad=\left\{
  \begin{array}{ll}
  e^{-\beta(H-N)} & \mbox{if  \;$\int{dx}\,\rho(x)=N$}\\
  0 & \mbox{otherwise}.
  \end{array}
  \right.
  \end{eqnarray}
The static solution \eref{static} has been thus verified, and supplementing Dean's discussion has been completed.

\section{Discussion: connection with the deterministic equation \eref{MT}}
Here we would like to first confirm the saddle-point path of the Fokker-Planck \eref{Fokkresult} from exploiting the WKB-like approach made in Ref. \cite{WKB}. Setting, similarly to the WKB approximation, that
  \begin{equation}
  \label{WKB}
  P(\{\rho\},t)\propto\exp\left[\,-\beta\Phi(\{\rho\},t)\,\right],
  \end{equation}
we obtain the Hamilton-Jacobi like equation:
  \begin{equation}
  \label{HJ}
  \fl \frac{\partial \Phi(\{\rho\},t)}{\partial t}
  =\int dx\,\frac{\delta \Phi(\{\rho\},t)}{\delta\rho(x)}
  \nabla\cdot L[\,\rho(x)\,]\nabla\,\left[
  \frac{\delta \Phi(\{\rho\},t)}{\delta\rho(x)}
  -\frac{\delta H(\{\rho\})}{\delta\rho(x)}\right].
  \end{equation}
A short-cut way of deriving from this the most probable (or saddle-point) path $\{\bar{\rho}\}$ is to expand $\Phi$ and $H$ around $\{\bar{\rho}\}$ as
  \begin{eqnarray}
  \label{expansion}
  \eqalign{
  \fl\Phi(\{\rho\},t)=\Phi(\{\bar{\rho}\},t)+\frac{1}{2}\int dx\,dy\;
  [\,\rho(x)-\bar{\rho}(x)\,]\,\Phi"(x-y)\,[\,\rho(y)-\bar{\rho}(y)\,]+\cdots\\
  \fl J(\{\rho\})=\left.J(\{\bar{\rho}\})+\frac{\delta J(\{\rho\})}
  {\delta\rho}\right|_{\{\rho\}=\{\bar{\rho}\}}\,[\,\rho(x)-\bar{\rho}(x)\,]+\cdots\>,
  }
  \end{eqnarray}
with $J(\{\rho\})\equiv L[\,\rho(x)\,]\nabla\delta H(\{\rho\})/\delta\rho(x)$. Substitution of these into equation \eref{HJ} yields in $\mathcal{O}[\,\rho(x)-\bar{\rho}(x)\,]$
  \begin{equation}
  \label{MTconfirm}
  \frac{\partial\bar{\rho}(\,x,t)}{\partial t}=\left.\nabla\cdot L[\,\bar{\rho}(x,t)\,]\,\nabla\frac{\delta H(\{\rho\})}{\delta\rho}\right|_{\{\rho\}=\{\bar{\rho}\}}.
\end{equation}
Since $\{\bar{\rho}\}$ is to be in accord with the noise averaged density $\{\rho_{av}\}$, equation \eref{MTconfirm} implies that $\langle\hat{\rho}(x,t)\hat{\rho}(y,t)\rangle=\rho_{av}(x,t)\rho_{av}(y,t)$ for the first term on the rhs of \eref{MT}, i.e., no spatial correlation of noise-averaging. In other words, the above derivation reveals that the noise-averaged equation \eref{MT} is not the first member of the dynamical BBGKY hierarchy unlike the proposal by MT, but is only the mean-field equation for the saddle-point path of $P(\{\rho\},t)$.

We have thus validated the stochastic density functional equations, which must be a powerful tool for the understanding of supercooled fluids and glasses, via proving the irrelevance of MT's objection to Dean's argument in three ways: (i) demonstrating that standard manipulations enable to replace with the c-number density field $\rho$ the corresponding operator variable $\hat{\rho}$ in the stochastic equation \eref{rhohat} derived by Dean, (ii) verifying the static solution \eref{static} of the Fokker-Planck equation for the density distribution functional with the help of the conditional grand canonical partition function, and (iii) pointing out that the noise averaged path satisfying the deterministic equation \eref{MT} merely corresponds to the saddle-point one.

The next problem is how to solve these dynamical density functional equations, stochastic or deterministic. In previous works \cite{MT, BagMuna, Fuchizaki}, the static density functional theory has been exploited as input, and some justifications have been also described by Kawasaki and MT \cite{Kawasaki, MT}. However, the present discussion does not support these; from our point of view, what to suppose for incorporating the static theory remains an open problem.\vspace{8pt}

\hspace{-20pt}We acknowledge the financial support from the Ministry of Education, Science, Culture, and Sports of Japan.



\section*{References}
\begin{thebibliography}{}

\bibitem{review}
For instance, see G\"oetze G and Sj\"ogren L 1992 {\itshape Rep. Prog. Phys.} {\bf 55} 241,
\nonum\hspace{4pt}a reference from the viewpoint of the mode coupling theory.

\bibitem{Dean}
Dean D S 1996 {\itshape J. Phys. A}: {\itshape Math. Gen.} {\bf 29} L613

\bibitem{Kawasaki}
Kawasaki K 1994 {\itshape Physica A} {\bf 208} 35; 1998 {\itshape J. Stat. Phys.} {\bf 93} 527

\bibitem{Munakata}
Munakata T 1989 {\itshape J. Phys. Soc. Jpn.} {\bf 58} 2434

\bibitem{others}
Kirkpatrick T R and Thirumalai D 1989 {\itshape J. Phys. A}: {\itshape Math. Gen.} {\bf 22} L149
\nonum\hspace{4pt}Tanaka H 1999 {\itshape J. Chem. Phys.} {\bf 111} 3163

\bibitem{nonlinear}
Das S P, Mazenko G F, Ramaswamy S and Toner J 1985 {\itshape Phys. Rev. A} {\bf 32} 3139

\bibitem{onuki}
See, for example, Yamamoto R and Onuki A 1998 {\itshape Phys. Rev. Lett.} {\bf 81} 4915.

\bibitem{MT}
Marconi U M B and Tarazona P 1999 {\itshape J. Chem. Phys.} {\bf 110} 8032; 2000 {\itshape J. Phys.: Condens. Matter} {\bf 12} A413

\bibitem{BagMuna}
Munakata T 1977 {\itshape J. Phys. Soc. Jpn.} {\bf 43} 1723
\nonum\hspace{4pt}Bagchi B 1987 {\itshape Physica A} {\bf 145} 273

\bibitem{Justin}
Zinn-Justin J 1996 {\itshape Quantum Field Theory and Critical Phenomena} (Oxford University: Oxford) Chap.4

\bibitem{WKB}
Kubo R, Matsuo K and Kitahara K 1973 {\itshape J. Stat. Phys.} {\bf 9} 51

\bibitem{BBGKY}
See, for example, Hanson J P and Mcdonald I R 1986 {\itshape Theory of Simple Liquids} (Academic Press, London)

\bibitem{frusawa}
Frusawa H and Hayakawa R 1999 {\itshape Phys. Rev. E} {\bf 60} R5048

\bibitem{Fuchizaki}
Fuchizaki K and Kawasaki K 1999 {\itshape Physica A} {\bf 266} 400

\end{thebibliography}
\end{document}